\documentstyle[sprocl]{article}

\input{psfig}

\def\Journal#1#2#3#4{{#1} {\bf #2}, #3 (#4)}

\def\PRL{\em Phys. Rev. Lett.}

\def\be{\begin{equation}}
\def\ee{\end{equation}}
\def\bea{\begin{eqnarray}}
\def\eea{\end{eqnarray}}

\newcommand{\bm}[1]{\mbox{\boldmath $ #1 $}}
\newlength{\minitwocolumn}
\def\ZPA{{\em Z. Phys.} A}
\def\ANP{\em Adv. Nucl. Phys.}
\def\PRC{{\em Phys. Rev.} C}

\begin{document}

\title{EFFECT OF VECTOR MESON MASS DECREASE\\ON SUPERFLUIDITY
IN NUCLEAR MATTER}

\author{TOMONORI TANIGAWA}

\address{Department of Physics, Kyushu University,\\
Fukuoka 812-8581, Japan\\E-mail: tomo2scp@mbox.nc.kyushu-u.ac.jp} 

\author{MASAYUKI MATSUZAKI}

\address{Department of Physics, Fukuoka University of Education,\\
Munakata, Fukuoka 811-4192, Japan\\E-mail: matsuza@fukuoka-edu.ac.jp}


\maketitle\abstracts{$^1S_0$ pairing in nuclear matter is
investigated by taking the hadron mass decrease into account via the
``In-Medium Bonn potential'' which was recently proposed by Rapp {\it et
al.} The resulting gap is significantly reduced in comparison with the
one obtained with the original Bonn-B potential and we ascertain that
the meson mass decrease is mainly responsible for this reduction.}

Superfluidity in nuclear matter is one of the important issues in
nuclear physics since it forms the foundation for nuclear structure
theory and has close relationship to neutron-star physics. 
To describe a nuclear many-body system, relativistic models are
attracting attention as well as non-relativistic ones. This is due to the
success of Quantum Hadrodynamics (QHD) originated with the study by Chin and
Walecka in 1970's.~\cite{serot}

We studied $^1S_0$ pairing in nuclear matter by means of QHD, which has
been succeeded in describing the various phenomena of nuclei, that is,
not only the saturation property of nuclear matter, but spherical,
deformed and rotating nuclei. Therefore improvement on the
description of pairing which is essential in open shell nuclei would
make QHD more reliable. To speak of neutron-star physics,
relativistic effect might be important due to high density which is a
few times larger than the normal nuclear matter density.\\

As is well known, QHD is an effective field theory of hadronic
degrees of freedom. The Lagrangian density for the present model is like
this:

\vspace*{-0.5cm}
\begin{eqnarray}
{\cal L}&=&\bar\psi(i\gamma_\mu\partial^\mu-M)\psi
 + {1\over2}(\partial_\mu\sigma)(\partial^\mu\sigma)
  -{1\over2}m_\sigma^2\sigma^2 \nonumber\\
 &-&{1\over4}\Omega_{\mu\nu}\Omega^{\mu\nu}+{1\over2}m_\omega^2\omega_\mu\omega^\mu
 - {1\over4}\bm{R}_{\mu\nu}\cdot\bm{R}^{\mu\nu}
  +{1\over2}m_\rho^2\bm{\rho}_\mu\cdot\bm{\rho}^\mu \\
 &-&g_\sigma\bar\psi\sigma\psi-g_\omega\bar\psi\gamma_\mu\omega^\mu\psi
  -g_\rho\bar\psi\gamma_\mu\bm{\tau}\cdot\bm{\rho}^\mu\psi
 - {1\over{3}}g_{2}\sigma^3-{1\over{4}}g_{3}\sigma^4 ,
\nonumber
\end{eqnarray}
\label{eq:lag}

\begin{center}
\begin{equation}
\Omega_{\mu\nu}=\partial_\mu\omega_\nu-\partial_\nu\omega_\mu ,
\quad
\bm{R}_{\mu\nu}=\partial_\mu\bm{\rho}_\nu
                  -\partial_\nu\bm{\rho}_\mu .
\end{equation}
\end{center}

\noindent
Here $\psi$ is the nucleon field, $\sigma$ stands for $\sigma$-boson
field, $\omega$ for $\omega$-meson field and $\rho$ for $\rho$-meson
field. Non-linear self-coupling terms for $\sigma$-boson, which are
crucial for a realistic description of nuclear properties within
Relativistic Mean Field Theory (RMFT), are also included.
Deriving the particle-particle (p-p) channel interaction in the
relativistic models is one of the current topics though practical
approach, namely, the use of a non-relativistic force ({\it e.g.} Gogny
force) in p-p channel, has been adopted in the finite nuclei
calculations.\\

The first study of pairing by means of QHD was done by Kucharek and Ring
in 1991.~\cite{kucharek} In order to incorporate the pairing field,
meson fields must be quantized and the anomalous Green's function is
defined by the Gor'kov factorization. The customary manipulation gives
Relativistic-Hartree-Bogoliubov (RHB) equation. Hence a transcendental
equation of effective nucleon mass for the mean field and an ordinary
non-linear gap equation for the pairing field organize the non-linear
system of equations.

The resulting gap obtained by adopting the one-boson-exchange (OBE)
interaction with RMFT parameter set such as NL1 is too large to achieve
a consensus.  Aforementioned ``practical approach'' for finite nuclei is
a remedy for this excessive gap problem. In nuclear matter,
alternatively, a realistic force can be used as p-p interaction and
produces a consistent result with the non-relativistic studies, where
the maximum pairing gap ranges from 2-4 MeV.~\cite{rummel}\\

A possible extension in alignment with the above policy is to take the
change of the hadron properties into consideration.  In the pioneering
work done by Brown and Rho, it was pointed out that the change of hadron
masses may occur in conformity with the chiral symmetry
arguments.~\cite{brownrho} Although there is still controversy on this
subject, some experiments seem to support the vector meson mass
decrease. The change is expressed by the linear relation between the
masses and the density:

\begin{equation}
  {{M^\ast}\over{M}}
={{m_{\rho,\omega}^\ast}\over{m_{\rho,\omega}}}
={{\Lambda_{\rho,\omega}^\ast}\over{\Lambda_{\rho,\omega}}}
=1-C{\rho\over{\rho_0}} , \quad C=0.15,
\end{equation}
\label{eq:brs}

\noindent
where $M$ is the nucleon mass, $m_{\rho, \omega}$ are the masses of the
$\rho$- and $\omega$-meson and $\Lambda_{\rho, \omega}$ are the cutoff
masses in the form factors applied to each nucleon-meson vertex. The
scaling factor $C$ is taken to be $0.15$, which is almost in line with
the one obtained from QCD sum rules. This relation is often referred to 
as ``Brown-Rho (BR) scaling.''

Then, Rapp {\it et al.} showed that hadron mass decrease conforming to
this scaling was compatible with the saturation property of nuclear
matter.~\cite{rapp} They constructed the OBE potential just by adding
two extra $\sigma$-bosons to the original Bonn-B potential with slightly
modified parameters.
This ``In-Medium Bonn potential'' makes the investigation of the medium
effects on superfluidity quite tractable. Accordingly we adopt this
potential as the p-p interaction in the gap equation.\\

\noindent
\begin{minipage}{0.45\textwidth}
\vspace*{-0.3cm}
  \setlength{\baselineskip}{3.5mm}
  \psfig{figure=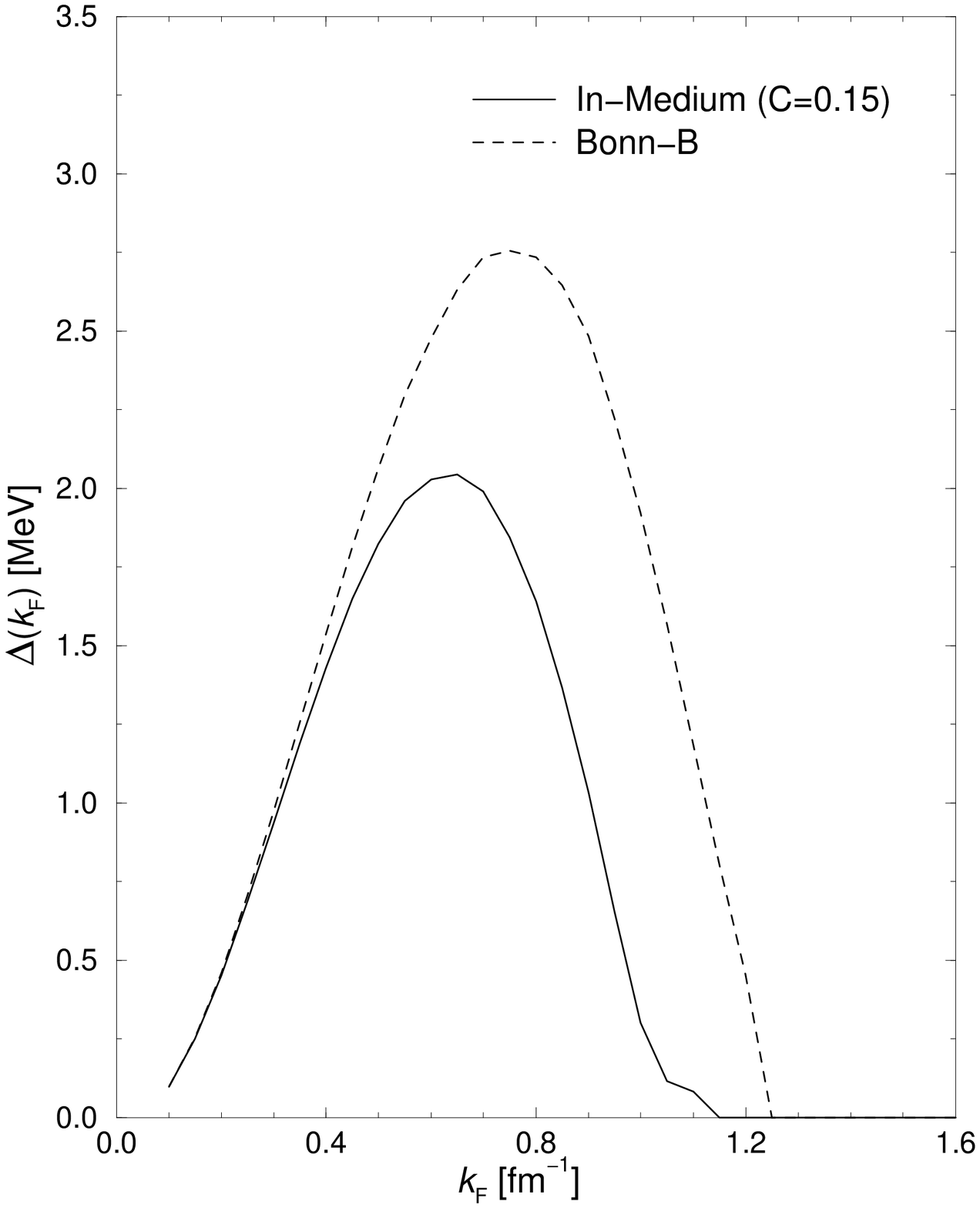,width=5.0cm,height=5.8cm}
\begin{footnotesize}
  Figure~\ref{fig:kfg2}: The pairing gaps at the Fermi surface as a function
 of Fermi momentum $k_{\rm F}$. Details are in the text.
\end{footnotesize}
  \refstepcounter{figure} 
  \label{fig:kfg2}
\end{minipage}
\vspace*{0.2cm}
\hspace*{0.03\textwidth}
\begin{minipage}{0.5\textwidth}
\vspace*{-0.3cm}
  \setlength{\baselineskip}{3.5mm}
  \psfig{figure=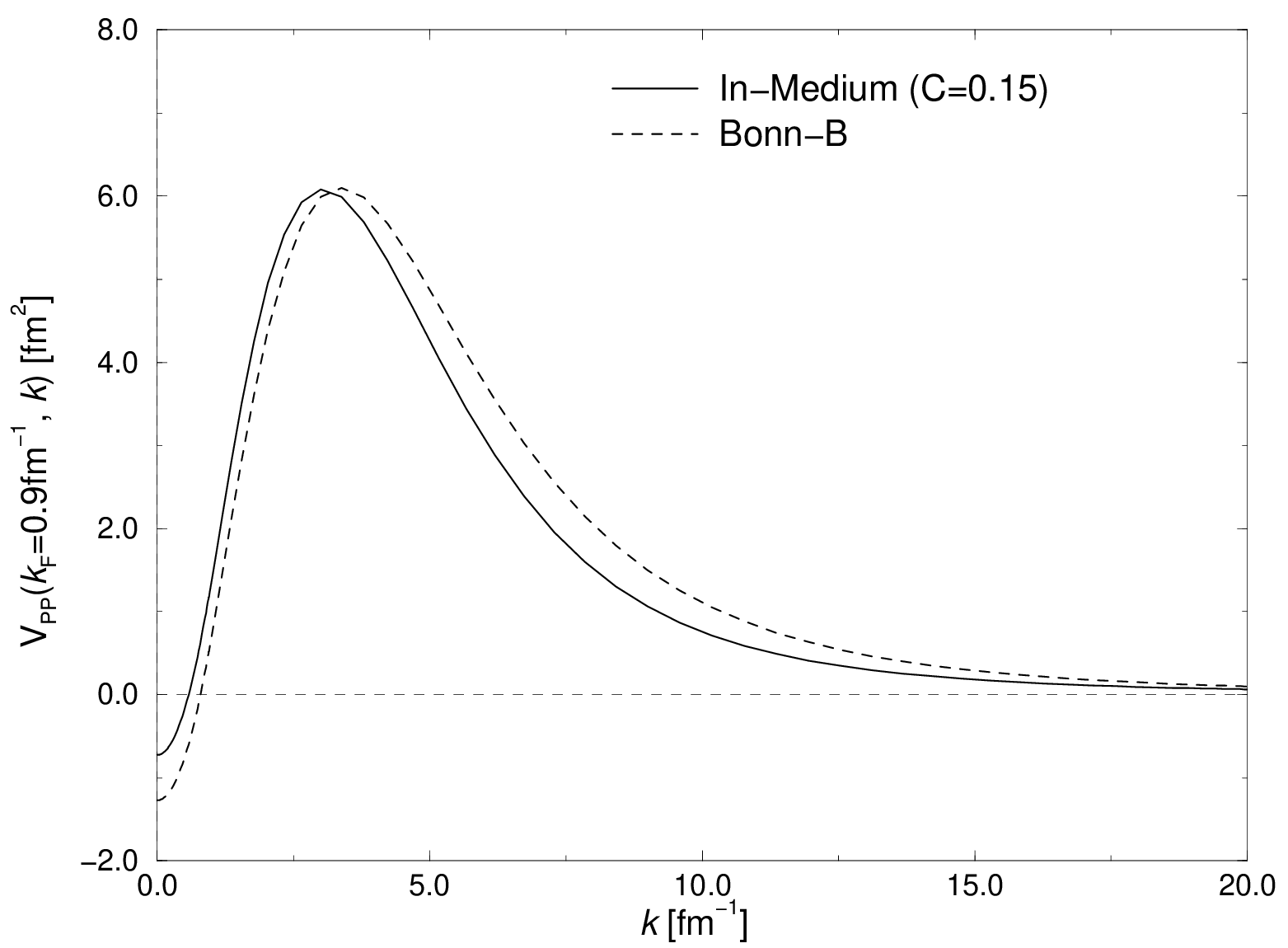,width=6.2cm,height=5cm}  
\begin{footnotesize}
  Figure~\ref{fig:pot2}: The particle-particle channel interactions as a
 function of momentum, with a Fermi momentum $k_{\rm F}=0.9$
 fm$^{-1}$. Details are in the text.
\end{footnotesize}
  \refstepcounter{figure} 
  \label{fig:pot2}
\end{minipage}

The resulting pairing gap at the Fermi surface is shown in
Figure~\ref{fig:kfg2}, which is drawn as a function of Fermi momentum
with the solid line. The dashed line corresponds to the gap obtained by
using the original Bonn-B potential. Comparison shows that the
inclusion of hadron mass decrease in concert with BR scaling
reduces the gap significantly, however, the values stay in the
physically acceptable range.

Before going into the detailed discussion of the gap, the structure of
the gap equation has to be reviewed to clarify how the contributions
come from each momentum region.
As pointed out by Rummel and Ring,~\cite{rummel} on the one hand,
positive contributions come from the low- and high-momentum region
where the gap has the opposite sign to the potential, which is due to
the minus sign in the integrand of the gap equation. On the other hand,
negative contributions come from the intermediate-momentum region where
the both potentials have a repulsive peak.

Next, to see the mass decrease effects on the p-p interaction, In-Medium
Bonn potential and the original Bonn-B potential are given in
Figure~\ref{fig:pot2}, which represents the shift of the In-Medium Bonn
potential to the lower-momentum region. In other words, the hadron mass
decrease leads to reduction of magnitude in the region giving the
positive contributions to the gap as mentioned above. This is the main
reason why reduction of the gap occurs in the case of In-Medium Bonn
potential.

\noindent
\begin{minipage}{0.55\textwidth}
\setlength{\parindent}{0.25in}
\vspace*{0.1cm}
But a question may arise: which hadron is responsible for this
reduction, nucleon or meson? In order to answer this, we also calculate
the gap applying BR scaling only to either hadron.
The result is shown in Figure~\ref{fig:kfg3}, where the solid, dashed
and long-dashed line correspond to the case of decreasing only the
nucleon mass, only the meson masses and the both, respectively. This
figure ascertains that the reduction of the gap is mainly due to the
meson mass decrease though the nucleon is responsible for it to some
extent.
\end{minipage}
\vspace*{0.1cm}
\hspace{0.03\textwidth}
\begin{minipage}{0.40\textwidth}
  \setlength{\baselineskip}{3.5mm}
\vspace*{-0.4cm}
  \psfig{figure=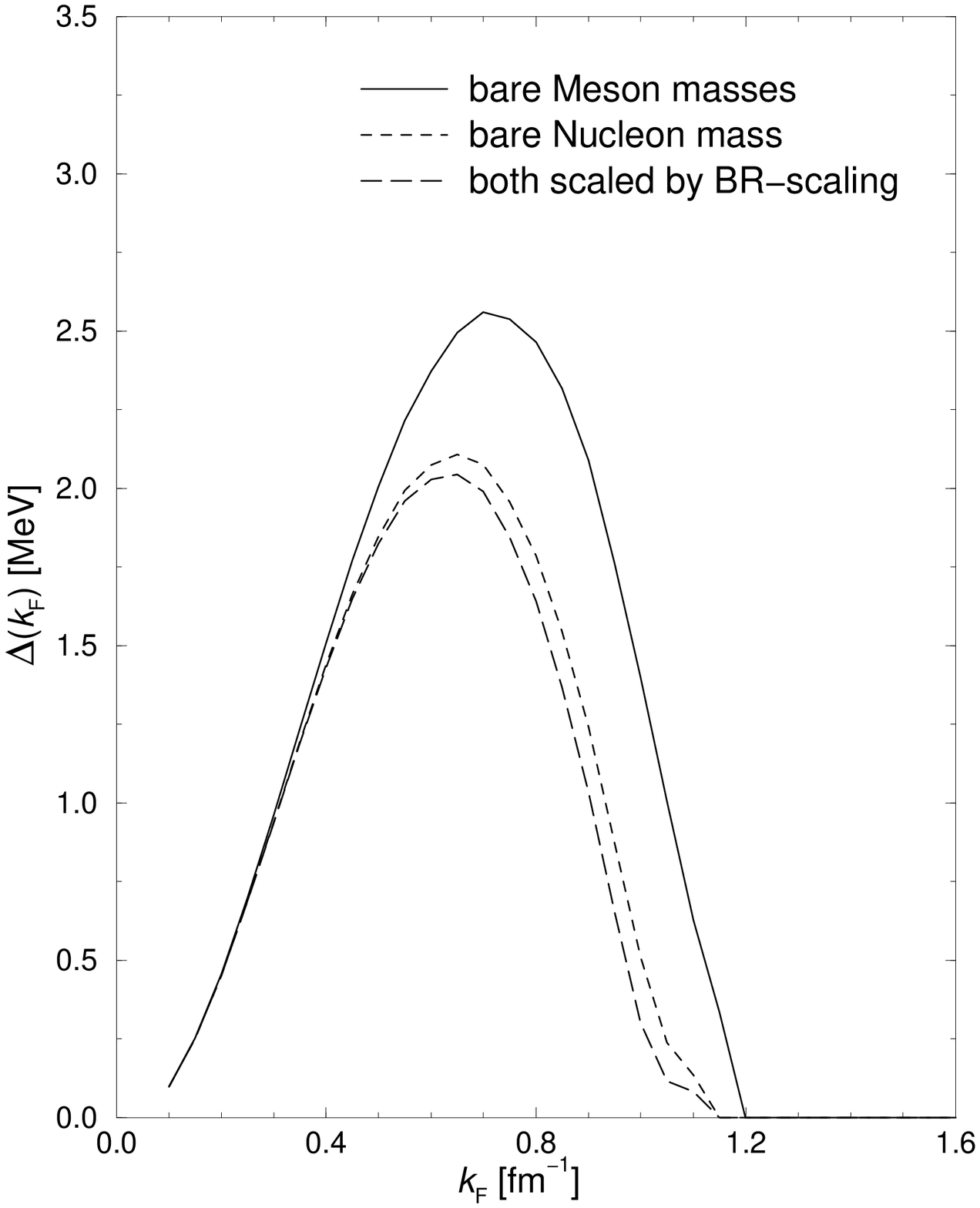,width=5.0cm,height=5.5cm}
 \begin{footnotesize}
  Figure~\ref{fig:kfg3}: The pairing gaps obtained with the partially BR
  scaled p-p interactions. Details are in the text.
 \end{footnotesize}
 \refstepcounter{figure} 
 \label{fig:kfg3}
\end{minipage}
\vspace*{0.1cm}

In summary, we performed the numerical calculation of the pairing gap
using In-Medium Bonn potential as the p-p interaction. The resulting gap
is considerably reduced in comparison with the one obtained with the
original Bonn-B potential and is consistent with the non-relativistic
studies. The use of the meson theoretic potential reveals that the
vector meson mass decrease, or lengthening the range of repulsive forces
accounts for the reduction.  Whether the polarization of Dirac sea as
well as Fermi sea is effective for it remains a conjecture left for
further investigation.~\cite{mm}

\end{document}